\documentclass[aps,pre,twocolumn,groupedaddress,superscriptaddress,showpacs]{revtex4-1}
\usepackage{epsfig,amssymb,amsmath,graphicx,subfigure,hyperref}
\usepackage{tabularx}
\usepackage{setspace}
\usepackage{color}
\usepackage{array} 
\newcommand{\PreserveBackslash}[1]{\let\temp=\\#1\let\\=\temp} \newcolumntype{C}[1]{>{\PreserveBackslash\centering}p{#1}} \newcolumntype{R}[1]{>{\PreserveBackslash\raggedleft}p{#1}} \newcolumntype{L}[1]{>{\PreserveBackslash\raggedright}p{#1}} 
\usepackage{graphicx}
\usepackage{dcolumn}
\usepackage{bm}
\usepackage{comment}
\hypersetup{hidelinks}

\begin{document}
\title{Configurational-space separation and structure selection in three hard squares}
\author{Yuheng Yang}
\affiliation{Key Laboratory of Artificial Micro- and Nano-structures of Ministry of Education and School of Physics and Technology, Wuhan University, Wuhan 430072, China}

\author{Meng Xiao}
\affiliation{Key Laboratory of Artificial Micro- and Nano-structures of Ministry of Education and School of Physics and Technology, Wuhan University, Wuhan 430072, China}
\affiliation{Wuhan Institute of Quantum Technology, Wuhan 430206, China}

\author{Duanduan Wan}
\email[E-mail: ]{ddwan@whu.edu.cn}
\affiliation{Key Laboratory of Artificial Micro- and Nano-structures of Ministry of Education and School of Physics and Technology, Wuhan University, Wuhan 430072, China}
\date{\today}

\begin{abstract}

Self-assembly of hard particles with diverse shapes gives rise to a rich variety of structures through excluded-volume constraints alone. Here we show that even a minimal system of three hard squares confined in a two-dimensional periodic box exhibits nontrivial configurational behavior relevant to structure selection. As the packing fraction increases, radial distribution functions obtained from Markov-chain Monte Carlo and uniform non-overlapping insertion sampling agree at low densities, deviate markedly over an intermediate range, and converge again at higher densities. Pressure measurements provide strong numerical evidence that the discrepancy originates from the separation of the allowed configurational space into two disconnected regions above a characteristic density. We identify the separation density as $\phi_{\rm sep}=3/5$, construct explicit overlap-free transition pathways connecting the two regions immediately below it, and quantify their relative configurational-space volumes. At higher packing fractions, an approximately L-shaped arrangement of the particle centers becomes strongly favored over a staggered one, revealing a structural motif characteristic of tetratic and square-lattice ordering in larger hard-square systems. These results show that excluded-volume geometry can govern both configurational connectivity and local structure selection even in a three-particle system, revealing how signatures of many-particle
self-assembly can already emerge in the few-particle limit.


\end{abstract}


\maketitle


\textit{Introduction.} Hard-particle systems---where particles interact solely through
excluded-volume constraints---exhibit a remarkable variety of
self-assembled structures determined by particle shape and packing
fraction (e.g., Refs.~\cite{Glotzer2007,Torquato2009,Haji-Akbari2009,Marechal2010,
Agarwal2011,Gang2011,Damasceno2012,Smallenburg2012,Ni2012,
Avendano2012,Gantapara2013,bernard2011,Anderson2017,
Lei2018_helix,Klotsa2018,Wan2019, Wan2021,Qin2023,Wan2023_chiral,
Li2025}), ranging from disordered fluids
\cite{wood1957,hoover1968} and liquid crystals
\cite{frenkel1987,mederos2014,lettinga2007} to plastic crystals
\cite{Ni2012,meijer2017,gantapara2015}, quasicrystals
\cite{Haji-Akbari2009,fayen2024}, and complex crystals with unit cells containing tens of particles \cite{Damasceno2012}.
Previous studies have mainly focused on many-particle systems, addressing collective ordering, equilibrium phases, and phase transitions (e.g., Refs.~\cite{Haji-Akbari2009,Marechal2010,Agarwal2011,Ni2012,
Avendano2012,Gantapara2013,bernard2011,Anderson2017,
Wan2019,meijer2017,gantapara2015,fayen2024}).
Large-scale simulations, for example, have shown that melting
scenarios depend sensitively on particle shape and may require systems
containing up to millions of particles to resolve the decay of
order-parameter correlations
\cite{bernard2011,Michael2013_pressure,russo2017,hou2024}.
By contrast, the few-particle limit remains much less explored,
particularly with respect to how the connectivity of the allowed
configurational space affects structural accessibility and selection,
and whether local ordering tendencies characteristic of many-particle
systems can already emerge at such minimal scales.

\begin{figure}
\includegraphics[width=\columnwidth]{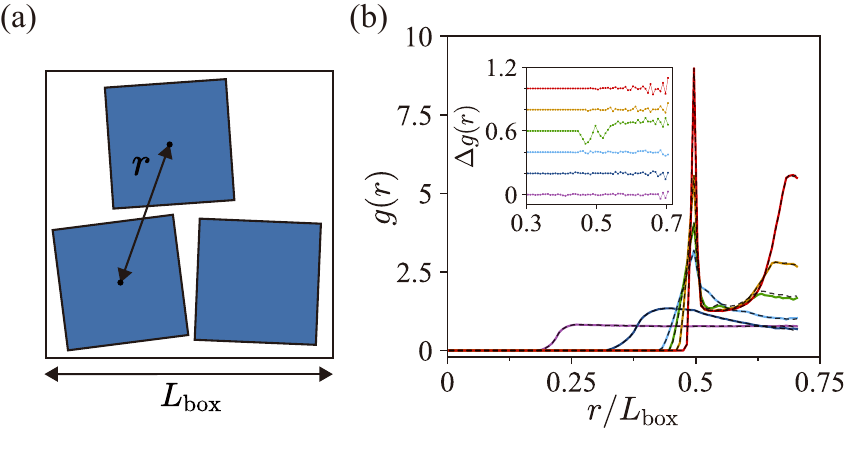} 
\caption{(a) Schematic illustration of three hard squares in a two-dimensional periodic box. The variable $r$ denotes the center-to-center distance between two squares. (b) Radial distribution function $g(r)$ for packing fractions $\phi = 0.1, 0.3, 0.55, 0.6, 0.65,$ and $0.7$ (increasing from bottom to top). Gray dashed lines show the results from uniform insertion sampling, while solid lines correspond to MCMC results. Inset: the difference $\Delta g(r)$ between the two methods; curves are vertically offset by 0.2 for clarity.}
\label{fig_rdf}
\end{figure}

To address these questions, we study three hard squares confined in a
two-dimensional periodic square box [Fig.~\ref{fig_rdf}(a)] using both
Markov-chain Monte Carlo (MCMC) and uniform non-overlapping insertion sampling. We find that the
two methods yield markedly different structural statistics near
$\phi\simeq0.6$. Analysis of the pressure provides strong numerical
evidence that the allowed configurational space separates into two
disconnected regions above a separation density $\phi_{\rm sep}$. Explicit overlap-free
transition pathways below this density reveal a narrow geometric
bottleneck, and a geometric construction yields
$\phi_{\rm sep}=3/5$. At higher packing fractions, we further find a
pronounced preference for an approximately L-shaped arrangement of the particle centers over a staggered one.
This local ordering motif is characteristic of both
tetratic and square-lattice ordering in larger hard-square systems.
These results show that even three hard squares can exhibit nontrivial
configurational connectivity and local ordering tendencies relevant to
many-particle self-assembly.

\begin{figure}
\includegraphics[width=\columnwidth]{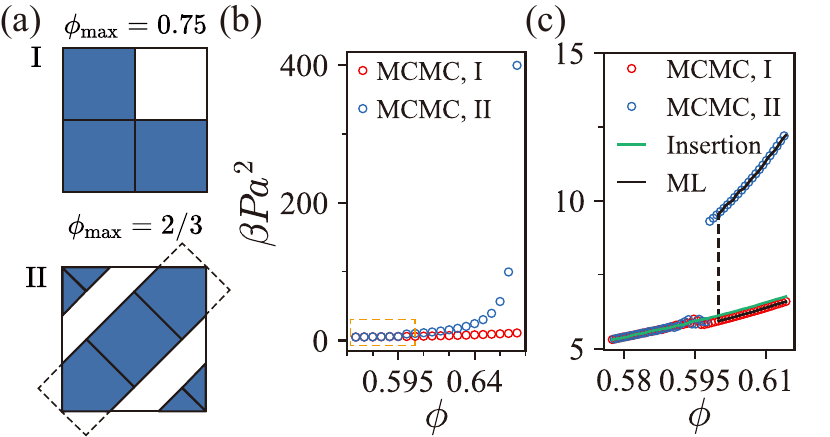} 
\caption{(a) Maximally packed configurations for two distinct configuration classes, I and II, with $\phi_{\max} = 0.75$ and $2/3$, respectively. (b) Pressure $\beta P a^2$ as a function of packing fraction $\phi$, obtained from MCMC simulations initialized from configuration classes I and II, respectively. (c) Zoomed-in view of the region highlighted by the orange dashed box in (b), with additional simulation points. The green line shows the results obtained from the uniform insertion method. The black lines indicate the two branches classified by machine learning. The vertical dashed black line marks $\phi = 0.6$. }
\label{fig_pressure}
\end{figure}

\textit{Methods.}
We study three hard squares confined in a square box with periodic
boundary conditions using the hard-particle Monte Carlo module of
HOOMD-blue \cite{Glaser2015,Anderson2020}. In MCMC sampling, the
particles evolve through local translational and rotational moves.
For the calculation of $g(r)$ in Fig.~\ref{fig_rdf}(b), the system is
initialized from the class-I configuration shown in
Fig.~\ref{fig_pressure}(a) and uniformly rescaled to the target packing
fraction. For the pressure calculations in Fig.~\ref{fig_pressure},
separate MCMC simulations are initialized from representative
configurations of classes I and II to obtain the two pressure branches.
Pressure data are reported in dimensionless form as $\beta P a^2$, where $\beta=(k_{\rm B}T)^{-1}$ and $a$ is the square side length; $\beta=1$ and $a=1$ are used in the simulations.
In uniform non-overlapping insertion sampling, three particles are
placed sequentially at random positions and orientations, and a
configuration is accepted only if all particles and their periodic
images are non-overlapping \cite{Wan2018,Wan2022}. Thus, this method
samples the allowed configurational space uniformly. Details of the
sampling protocols, pressure calculations, machine-learning
classification, and Monte Carlo chain-of-states construction are
provided in the Supplemental Material.

\textit{Results and discussion.}
Figure~1(b) compares the radial distribution function $g(r)$ obtained
from MCMC and uniform insertion sampling at selected packing fractions,
where $r$ denotes the center-to-center distance between square particles [Fig.~\ref{fig_rdf}(a)].
For the MCMC simulations, the system is initialized from the L-shaped
class-I configuration rescaled to the target packing fraction.
The two methods agree well for $\phi\leq0.55$, as indicated by the small
$\Delta g(r)$ values shown in the inset. At $\phi=0.6$, however, a clear
discrepancy emerges, with $\Delta g(r)$ becoming significantly larger.
At higher packing fractions, $\phi=0.65$ and $0.7$, the difference
decreases again.

To understand this anomalous behavior, we further examine the pressure of the system. 
The MCMC pressures are shown by the red circles in
Fig.~\ref{fig_pressure}(b)
(see Supplemental Material, Sec.~III, for details). Overall, the pressure increases with $\phi$, while any anomalous behavior near $\phi \simeq 0.6$ is not clearly resolved on this scale.
We therefore perform additional simulations in the region highlighted by the orange dashed box in Fig.~\ref{fig_pressure}(b), as shown in the enlarged view in Fig.~\ref{fig_pressure}(c).
The denser sampling reveals a small range around
$\phi\simeq0.595$ in which the MCMC pressure varies irregularly and
shows substantial run-to-run variation. In contrast, the pressure
obtained from uniform insertion sampling varies smoothly across this
range [green solid line in Fig.~\ref{fig_pressure}(c)]. Below this
irregular range, the insertion and MCMC pressures agree closely,
whereas on the higher-$\phi$ side, the insertion pressure lies
systematically above the MCMC values shown by the red circles.

As the pressure obtained from uniform insertion sampling represents an
ensemble average over all allowed configurations, its deviation from
the MCMC results indicates that the MCMC trajectories do not fully
explore the allowed configurational space within the finite simulation
time. By examining the configurations sampled by the two methods, we
identify a second configuration class, denoted class II, in which the
three squares form a tilted linear arrangement at $45^\circ$ with
respect to the box sides [Fig.~\ref{fig_pressure}(a), II]. The maximum
packing fraction of class II is
$\phi_{\max}^{\rm II}=2/3$, lower than
$\phi_{\max}^{\rm I}=0.75$ for class I~\cite{Blair2012}. We then perform MCMC simulations
initialized from class-II configurations.
The resulting pressures, shown by the blue circles in Fig.~\ref{fig_pressure}(b), increase rapidly as $\phi$ approaches $2/3$.
A closer examination of the region near $\phi\simeq0.6$ shows that
two distinct pressure branches emerge for $\phi\gtrsim0.6$
[Fig.~\ref{fig_pressure}(c)]. The class-II branch lies above the
class-I branch, while the uniform insertion results lie between the
two. For $\phi\lesssim0.595$, the three results converge to the same
curve. These observations provide strong numerical evidence that the
allowed configurational space separates into two disconnected regions
above $\phi\simeq0.6$ and becomes connected below this density. The
irregular MCMC pressures around $\phi\simeq0.595$ can then be
understood as a finite-time sampling regime in which different
trajectories access the class-II region to different extents, leading
to run-dependent pressure estimates.

\begin{figure*}
   \centering
   \includegraphics[width=18cm]{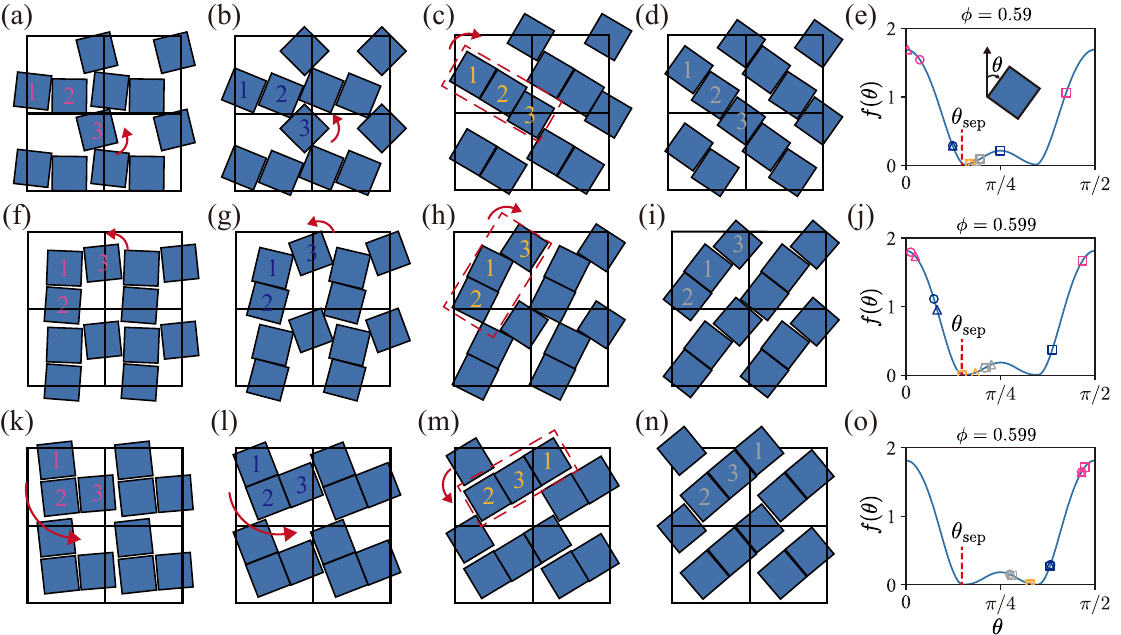}
   \caption{
Transition pathways between the two configuration classes.
(a)--(d) Representative configurations along an MCMC transition from the larger configuration-space region associated with class I to the smaller region associated with class II at $\phi=0.59$.
In (a) and (b), the red arrows indicate the subsequent rotation of
square 3.
In (c), the three squares have nearly the same orientation, as highlighted by the dashed red rectangle; the red arrow indicates their subsequent collective rotation.
(e) Area-normalized probability density $f(\theta)$ of the square orientation angle $\theta$ at $\phi=0.59$, obtained from uniform insertion sampling and normalized such that
$\int_{0}^{\pi/2} f(\theta)\,d\theta=1$.
Here, $\theta$ is defined as the smallest clockwise rotation of a fixed laboratory axis required to align it with an edge of the square, as illustrated in the inset.
Circles, triangles, and squares denote particles 1, 2, and 3, respectively.
The symbol colors correspond to the configurations in (a)--(d) and match the colors of the corresponding particle labels.
The vertical red dashed lines mark the separation angle
$\theta_{\rm sep}=\tan^{-1}(1/2)$ (see Supplemental Material, Sec.~V).
(f)--(i) and (k)--(n) Two transition pathways constructed using the Monte Carlo chain-of-states method at $\phi=0.599$.
In (f) and (g), the red arrows indicate the subsequent rotation of square 3.
In (h), the three squares have nearly the same orientation, as highlighted by the dashed red rectangle; the red arrow indicates their subsequent collective rotation along the chain.
In (k) and (l), the red arrows indicate the subsequent collective rotation of the three squares.
(j) and (o) Same as (e), but at $\phi=0.599$, with symbols indicating
the particle orientations corresponding to configurations (f)--(i)
and (k)--(n), respectively.
}
\label{fig_transition}
\end{figure*}

We next examine how the two configuration classes become connected near $\phi\simeq0.6$. 
Representative snapshots along a transition pathway observed directly by MCMC at $\phi=0.59$ are shown in Fig.~\ref{fig_transition}(a)--(d) (see also Supplemental Movie~S1). 
The pathway starts from a configuration in the larger configurational region associated with class I [Fig.~\ref{fig_transition}(a)], with the particle orientations indicated by the pink symbols in Fig.~\ref{fig_transition}(e). 
Squares 1 and 2 then rotate clockwise, while square 3 rotates counterclockwise, as indicated by the red arrow, leading to the intermediate configuration in Fig.~\ref{fig_transition}(b). 
The corresponding orientations are marked by the dark-blue symbols in Fig.~\ref{fig_transition}(e). 
Further rotation brings square 3 into a nearly collinear arrangement with squares 1 and 2 [Fig.~\ref{fig_transition}(c)], as highlighted by the dashed red rectangle. 
At this stage, the three squares also have nearly the same orientation,
with their orientations lying in a low-probability region of the
orientation distribution in Fig.~\ref{fig_transition}(e), indicating
that the system has reached a narrow bottleneck in configurational
space.
The three squares subsequently undergo a collective clockwise rotation, as indicated by the red arrow in Fig.~\ref{fig_transition}(c), and evolve into the configuration shown in Fig.~\ref{fig_transition}(d), thereby passing through the bottleneck and entering the smaller configurational region associated with class II.

As $\phi$ approaches $0.6$, the connecting channel becomes increasingly narrow, making such transitions difficult to sample directly by MCMC. 
We therefore employ a Monte Carlo chain-of-states method in configurational space to search for transition pathways closer to this density. 
In this approach, the two endpoints of each chain are fixed to representative configurations of classes I and II, respectively, while the intermediate configurations are relaxed subject to the hard-particle constraints (see Supplemental Material, Sec.~IV, for details). 
Remarkably, even at $\phi=0.599$, the method identifies two distinct overlap-free pathways connecting classes I and II, as shown in Fig.~\ref{fig_transition}(f)--(i) and (k)--(n); see also Supplemental Movies~S2 and S3.
The pathway in Fig.~\ref{fig_transition}(f)--(i) closely resembles the transition observed directly by MCMC at $\phi=0.59$ in Fig.~\ref{fig_transition}(a)--(d). 
By contrast, along the second pathway
[Fig.~\ref{fig_transition}(k)--(n)], the three squares undergo a
collective counterclockwise rotation and pass through a nearly
collinear bottleneck, highlighted by the red dashed rectangle in
Fig.~\ref{fig_transition}(m), before entering the class-II region.
Despite their different rotational routes, all three pathways pass
through essentially the same bottleneck, characterized by a compact,
tilted arrangement of three nearly parallel squares.
This recurring geometric structure suggests an origin for the
configurational-space separation near $\phi\simeq0.6$. As shown in the
Supplemental Material, Sec.~V, a geometric construction
based on this bottleneck yields
$\phi_{\rm sep}=3/5=0.6$ and
$\theta_{\rm sep}=\tan^{-1}(1/2)$, with the latter indicated by the red
dashed lines in Fig.~\ref{fig_transition}. These values are consistent
with the separation density and bottleneck orientation observed in the
simulations.

When the two configurational regions are disconnected, their relative
volumes can be inferred from three pressures: the pressures of the two
individual branches and the ensemble-averaged pressure obtained from
uniform insertion sampling. At the separation density $\phi_{\rm sep}=0.6$, this analysis gives $V_2/V_{\mathrm{tot}}\approx0.0508$, where $V_2$ denotes the
configurational-space volume associated with class II and
$V_{\mathrm{tot}}$ denotes the total configurational-space volume
(see Supplemental Material, Sec.~III).
To test the consistency of this two-region picture, for $\phi \geq 0.6$
we further classify the successful configurations generated by uniform
insertion sampling into classes I and II using a machine-learning
classifier trained on MCMC samples, and compute the pressure of each
class separately (see Supplemental Material, Sec.~VI).
The resulting class-resolved pressures, shown by the black solid lines
in Fig.~\ref{fig_pressure}(c), agree well with the corresponding MCMC
branches. Using these class-resolved pressures together with the
ensemble-averaged insertion pressure at $\phi_{\rm sep}=0.6$ gives
$V_2/V_{\mathrm{tot}} \approx 0.0503$,
in close agreement with the branch-pressure estimate above.
This agreement further supports the
self-consistency of the two-region picture.

This configurational-space picture also explains the nonmonotonic
difference between the MCMC and uniform-insertion results in
Fig.~\ref{fig_rdf}(b). For $\phi\leq0.55$, the allowed configurational
space is connected, and the $g(r)$ curves obtained from the two sampling
methods agree closely. At $\phi=0.6$, the connecting channel has
narrowed to a marginal bottleneck. Consequently, the MCMC simulations
used for Fig.~\ref{fig_rdf}(b), which are initialized in class I,
remain effectively confined to the class-I region over the finite
simulation time, whereas uniform insertion samples both configurational
regions. This leads to the largest discrepancy between the two
$g(r)$ curves. As $\phi$ increases further to $0.65$, the relative
configurational-space volume of class II decreases, reducing its
contribution to the uniform-insertion ensemble and hence the difference
between the two methods. Finally, for
$\phi>\phi_{\max}^{\rm II}=2/3$, the class-II region no longer exists.
Thus, at $\phi=0.7$, both methods sample only the class-I region, and
their $g(r)$ results again show close agreement.


\begin{figure}
\includegraphics[width=\columnwidth]{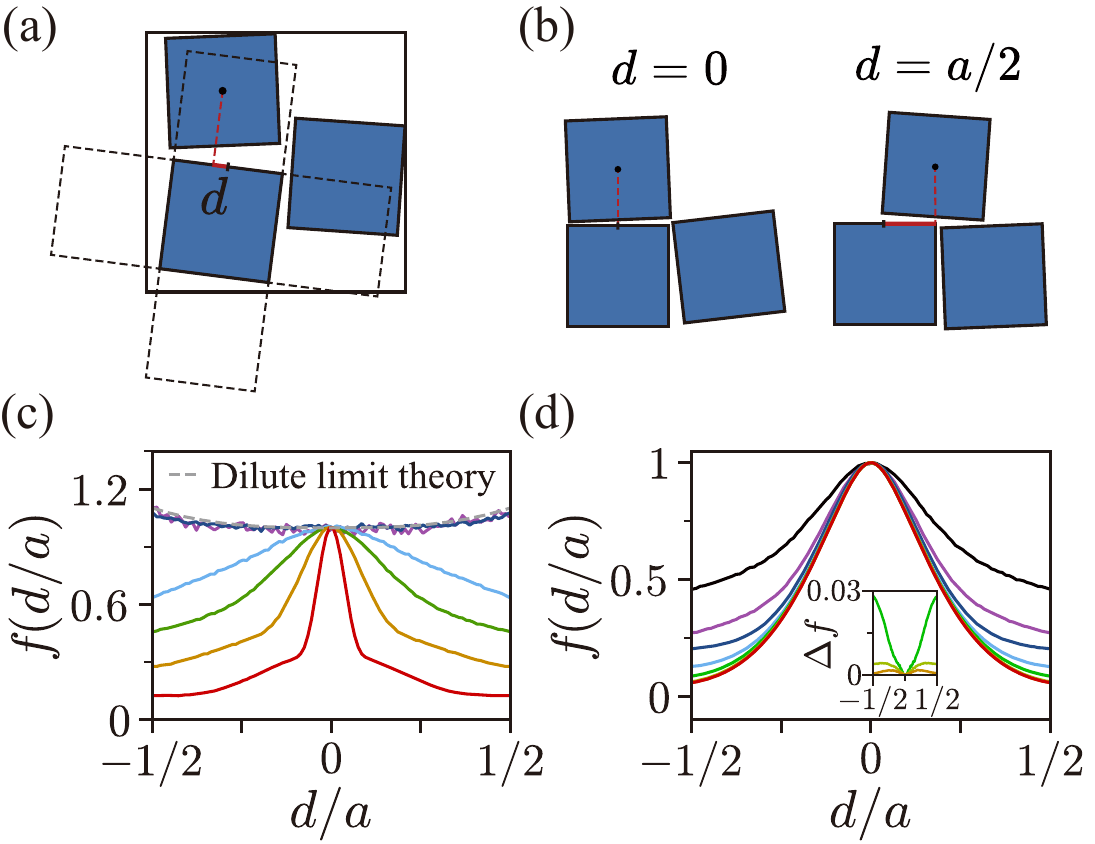} 
\caption{
Definition and distributions of the local order parameter $d/a$.
(a) Definition of $d$. For a reference square, we consider its four adjacent square regions, indicated by dashed outlines. If the center of mass of another square or one of its periodic images lies within one of these regions, it is projected perpendicularly onto the corresponding side of the reference square. The signed distance between the projection point and the midpoint of
that side is defined as $d$ (red segment), with the positive direction
chosen to follow the clockwise orientation around the reference square.
(b) Representative configurations with $d=0$ and $d=a/2$.
(c) Normalized frequency $f(d/a)$ as a function of $d/a$, with $f(0)$ normalized to unity. 
Away from the common normalized point at $d/a=0$, the curves from top to bottom correspond to packing fractions $\phi=0.1$, $0.3$, $0.55$, $0.6$, $0.65$, and $0.7$, respectively, using the same color scheme as in Fig.~\ref{fig_rdf}(b). 
The gray dashed line shows the dilute-limit theoretical prediction
(see Supplemental Material, Sec.~VII).
(d) $f(d/a)$ for systems with different particle numbers. For $N\geq 4$, the packing fraction is $\phi=0.8$; for $N=3$, the black curve is identical to the green curve at $\phi=0.6$ in (c), corresponding to an effective four-particle packing fraction $\phi_{\mathrm{eff}}=0.8$. Away from the common normalized point at $d/a=0$, the curves from top to bottom correspond to $N=3$, $4$, $9$, $36$, $100$, $400$, $900$, and $2500$. The inset shows $\Delta f=f_N(d/a)-f_{2500}(d/a)$ for $N=100$, $400$, and $900$.
}
\label{fig_order_parameter}
\end{figure}

To characterize the relative positions of neighboring squares, we
define a local order parameter as follows. As shown in
Fig.~\ref{fig_order_parameter}(a), for each reference square we
consider the four adjacent regions sharing one side with it, indicated
by the dashed outlines. If the center of mass of another square or one
of its periodic images lies within one of these regions, it is
projected perpendicularly onto the corresponding side of the reference
square. The signed distance between the projection point and the
midpoint of that side is defined as $d$, with the positive direction
chosen to follow the clockwise orientation around the reference square.
Normalizing by the square side length $a$, we use $d/a$ as the local
order parameter, which lies in the range $[-1/2,1/2]$. Figure~\ref{fig_order_parameter}(b) shows two representative local
arrangements with $d=0$ and $d=a/2$, both of which are accessible at
$\phi_{\max}^{\rm I}=0.75$. For $d=0$, the center of the third square
is aligned with that of one of the lower squares, giving the three
particle centers an approximately L-shaped arrangement. For $d=a/2$,
the third square is positioned above and between the two lower squares,
forming a staggered arrangement.

Figure~\ref{fig_order_parameter}(c) shows the normalized frequency
distribution $f(d/a)$ at several packing fractions, using the same
color scheme as in Fig.~\ref{fig_rdf}(b). Each distribution is
normalized such that $f(0)=1$. For each configuration, the three
squares are taken in turn as the reference square, and all other
squares and their periodic images satisfying the construction in
Fig.~\ref{fig_order_parameter}(a) contribute to the statistics.
At low packing fractions, $\phi=0.1$ and $0.3$, the distribution has
a shallow minimum at $d/a=0$ and increases toward $d/a=\pm1/2$,
where it reaches its maximum. This weak variation is consistent with
the dilute-limit result shown by the gray dashed line in
Fig.~\ref{fig_order_parameter}(c)
(see Supplemental Material, Sec.~VII, for the derivation).
As $\phi$ increases, this trend reverses: configurations with $d/a=0$
become increasingly favored, whereas those near $d/a=\pm1/2$ are
progressively suppressed. The evolution of $f(d/a)$ therefore
indicates an increasing preference for the approximately L-shaped
local arrangement.
The same tendency is also evident in the high-density MCMC trajectory
shown in Supplemental Movie~S4, where approximately L-shaped
arrangements with $d=0$ are observed much more frequently than
staggered arrangements with $d=a/2$.

In Fig.~\ref{fig_order_parameter}(d), we compare $f(d/a)$ for systems
with different particle numbers. The black curve for $N=3$ is
identical to the curve at $\phi=0.6$ in
Fig.~\ref{fig_order_parameter}(c). At this density, the box area is
$L^2=5a^2$; placing four particles in the same box would correspond to
an effective packing fraction
$\phi_{\rm eff}=4a^2/L^2=0.8$. This provides a natural density
correspondence between the three-particle system at $\phi=0.6$ and
larger systems at $\phi=0.8$.
As $N$ increases, the distribution becomes increasingly concentrated
around $d/a=0$, indicating an enhanced preference for the same local
geometric arrangement. The inset shows
$\Delta f=f_N(d/a)-f_{2500}(d/a)$ for $N=100$, $400$, and $900$,
demonstrating that the difference from the $N=2500$ distribution
decreases with increasing $N$.
For large hard-square systems, $\phi=0.8$ lies in the tetratic regime
\cite{Anderson2017}. Under the same density
correspondence, higher packing fractions of the three-particle system
map to higher packing fractions of the many-particle system, where
square-lattice order emerges \cite{Anderson2017, Wojciechowski2004}. Consistently, the preference for
$d/a=0$ becomes more pronounced with increasing $\phi$ in the
three-particle system [Fig.~\ref{fig_order_parameter}(c)]. These
results indicate that the local structural preference associated with
both tetratic and square-lattice ordering is already present in the
three-particle system and becomes increasingly pronounced with system
size and packing fraction.

\textit{Conclusion.}
In conclusion, we have shown that the self-assembly of even three hard
squares in a periodic box exhibits rich configurational behavior
governed solely by excluded-volume constraints. A marked discrepancy
between MCMC and uniform insertion sampling near $\phi\simeq0.6$ led
us to identify a second configuration class. MCMC simulations
initialized from the two classes then reveal two distinct pressure
branches, while the uniform-insertion pressure lies between them,
providing strong numerical evidence for configurational-space
separation near $\phi\simeq0.6$. Explicit overlap-free transition
pathways below this density reveal a narrow geometric bottleneck
connecting the two configurational regions, and a geometric
construction based on this bottleneck yields the separation density
$\phi_{\rm sep}=3/5$. The relative configurational-space volumes of the two regions can
further be quantified from the MCMC branch pressures and the
ensemble-averaged insertion pressure.

At higher packing fractions, excluded-volume constraints also produce
a pronounced local structure selection, favoring an approximately
L-shaped arrangement over a staggered one. The corresponding local
ordering motif is characteristic of tetratic and square-lattice
ordering in larger hard-square systems. Our results therefore
illustrate, in a minimal three-particle setting, how excluded-volume
geometry can constrain configurational connectivity and select local
structures relevant to many-particle self-assembly.


\begin{acknowledgments}
This work was supported by the National Natural Science Foundation of China: Grant No.~12274330, Grant No.~12334015, and Grant No.~12274332. This work was also supported by the Young Top-Notch Talent Cultivation Program of Hubei Province. D.W. acknowledges the ``Xiaomi Young Scholar Program'' at Wuhan University.
\end{acknowledgments}

\bibliography{ref}
\end{document}


\title{Supplemental Material for ``Configurational-space separation and structure selection in three hard squares''}

\author{Yuheng Yang}
\affiliation{Key Laboratory of Artificial Micro- and Nano-structures of Ministry of Education and School of Physics and Technology, Wuhan University, Wuhan 430072, China}

\author{Meng Xiao}
\affiliation{Key Laboratory of Artificial Micro- and Nano-structures of Ministry of Education and School of Physics and Technology, Wuhan University, Wuhan 430072, China}
\affiliation{Wuhan Institute of Quantum Technology, Wuhan 430206, China}

\author{Duanduan Wan}
\email[E-mail: ]{ddwan@whu.edu.cn}
\affiliation{Key Laboratory of Artificial Micro- and Nano-structures of Ministry of Education and School of Physics and Technology, Wuhan University, Wuhan 430072, China}

\date{\today}
\maketitle

\vspace{1em}

\noindent
This PDF includes:

\begin{itemize}
    \item Captions for Supplemental Movies S1 to S4

    \item Simulation details

    \item Pressure calculation and configurational-space volume

    \item Monte Carlo chain-of-states construction of transition paths

    \item Geometric origin of the configurational-space separation density

    \item Classification of high-density configurations

    \item Order parameter in the dilute limit
\end{itemize}

\vspace{0.5em}

\noindent
Other supplemental materials for this manuscript include the following:

\begin{itemize}
    \item Supplemental Movies S1 to S4 (.mp4)
\end{itemize}

\clearpage

\section{Captions for Supplemental Movies S1--S4}
\label{sec:sm_movie}

\textbf{Movie S1:}
An MCMC transition trajectory connecting the two configuration classes
at $\phi=0.59$. Representative frames from this trajectory are shown
in Fig.~3(a)--(d) of the main text.

\textbf{Movie S2:}
A transition trajectory constructed using the Monte Carlo
chain-of-states method at $\phi=0.599$. Representative frames are shown
in Fig.~3(f)--(i) of the main text.

\textbf{Movie S3:}
A second transition trajectory constructed using the Monte Carlo
chain-of-states method at $\phi=0.599$. Representative frames are shown
in Fig.~3(k)--(n) of the main text.

\textbf{Movie S4:}
An MCMC trajectory at $\phi=0.74$, illustrating the stronger preference
for an approximately L-shaped arrangement of the particle centers over the staggered arrangement.

\section{Simulation details}
\label{sec:sm_simulation}
We perform hard-particle simulations and pressure calculations using the hard-particle Monte Carlo module of HOOMD-blue \cite{Glaser2015, Anderson2020}. The system consists of three hard squares of side length $a$ confined in a square box with periodic boundary conditions. For MCMC sampling, one MC step consists of attempting a random translational and rotational move for each particle. At each packing fraction, the translational and rotational step sizes are adjusted to maintain an acceptance rate of approximately 0.3. For the calculation of the radial distribution function $g(r)$ in Fig.~1(b) of the main text, the system is initialized from the class-I configuration shown in Fig.~2(a) of the main text and is uniformly rescaled with the box size to the target packing fraction. We perform $10^{10}$ MC steps and record one configuration every $10^{4}$ steps, yielding $10^{6}$ samples. For uniform insertion sampling, three particles are placed sequentially at random positions and orientations \cite{Wan2018, Wan2022}. If a trial particle overlaps with any previously placed particle or its periodic images, the entire configuration is rejected and the procedure is restarted. A configuration is accepted only when all three particles have been placed without overlap. For each packing fraction, this procedure is repeated until $10^{6}$ valid configurations are obtained. Thus, the two sampling protocols generate the same number of configurations at each packing fraction.

\section{Pressure calculation and configurational-space volume}
\label{sec:sm_volume}

We use the HPMC module in HOOMD-blue to calculate the pressure of the
hard-particle system~\cite{Anderson2016}. The pressure is evaluated
using the volume-perturbation technique~\cite{Eppenga1984,Brumby2011}.
Further details of the pressure calculation can be found in
Ref.~\cite{Yang2025}. 
For the pressure calculations shown in Fig.~2 of the main text, each
MCMC simulation consisted of $5\times10^9$ MC steps, with one
configuration sampled every five steps, yielding $10^9$ sampled
configurations for each pressure data point. The values $\beta P_1 a^2$ and
$\beta P_2 a^2$ quoted below were obtained from MCMC simulations initialized
from classes I and II, respectively. For uniform insertion sampling,
$10^7$ statistically independent non-overlapping configurations were
generated at each packing fraction and used to evaluate
$P_{\mathrm{tot}}$.

For $\phi\gtrsim0.6$, the MCMC trajectories initialized from classes I and II remain confined to two distinct configurational regions, giving rise to the two pressure branches observed in Fig.~2(c) of the main text. For a hard-particle system, the potential energy is zero for all allowed configurations and infinite for forbidden configurations. The Boltzmann factor is therefore unity within the allowed region of configuration space and zero otherwise. Consequently, the configurational partition function is proportional to the volume of the allowed configuration space. The relative volumes of the two disconnected regions can thus be inferred from the pressures of the two regions and the pressure of the full configuration space.

The pressures of the two disconnected regions associated with classes I and II are
\begin{equation}
    P_1
    = -\frac{\partial F_1}{\partial A_\mathrm{box}}
    = k_B T \frac{\partial \ln Q_1}{\partial A_\mathrm{box}}
    = \frac{k_B T}{Q_1}\frac{\partial Q_1}{\partial A_\mathrm{box}},
\label{Eq:press_1}
\end{equation}
and
\begin{equation}
    P_2
    = -\frac{\partial F_2}{\partial A_\mathrm{box}}
    = k_B T \frac{\partial \ln Q_2}{\partial A_\mathrm{box}}
    = \frac{k_B T}{Q_2}\frac{\partial Q_2}{\partial A_\mathrm{box}}.
\label{Eq:press_2}
\end{equation}
The pressure of the full configuration space is
\begin{equation}
    P_{\mathrm{tot}}
    = -\frac{\partial F_{\mathrm{tot}}}{\partial A_\mathrm{box}}
    = k_B T \frac{\partial \ln Q_{\mathrm{tot}}}{\partial A_\mathrm{box}}
    = \frac{k_B T}{Q_{\mathrm{tot}}}
      \frac{\partial Q_{\mathrm{tot}}}{\partial A_\mathrm{box}}.
\label{Eq:press_tot}
\end{equation}
Here $F_{1,2}$ and $Q_{1,2}$ are the free energies and configurational partition functions of the two disconnected regions, respectively, while $F_{\mathrm{tot}}$ and $Q_{\mathrm{tot}}$ are those of the full configuration space. Substituting Eqs.~\eqref{Eq:press_1} and \eqref{Eq:press_2} into Eq.~\eqref{Eq:press_tot}, we obtain
\begin{equation}
P_1 Q_1 + P_2 Q_2 = P_{\mathrm{tot}} Q_{\mathrm{tot}} .
\end{equation}
Using $Q_{\mathrm{tot}} = Q_1 + Q_2$, the relative weight of region II in the full configuration space is
\begin{equation}
    \frac{Q_2}{Q_{\mathrm{tot}}}
    =
    \frac{P_{\mathrm{tot}} - P_1}{P_2 - P_1}.
\end{equation}
At $\phi = 0.6$, the dimensionless pressures of the two regions are $\beta P_1 a^2 = 5.936$ and $\beta P_2 a^2 = 9.577$, while the pressure calculated from the uniform insertion method is $\beta P_{\mathrm{tot}} a^2 = 6.121$. Let $V_1$, $V_2$, and $V_{\mathrm{tot}}$ denote the corresponding
configurational-space volumes. We then obtain
\begin{equation}
  \frac{V_2}{V_{\mathrm{tot}}} = \frac{Q_2}{Q_{\mathrm{tot}}}
    =
    \frac{6.121 - 5.936}{9.577 - 5.936}
    \approx 0.0508 .
\end{equation}
Thus, at this packing fraction, the configuration-space region associated with class II accounts for approximately $5.1\%$ of the total configuration-space volume.

\section{Monte Carlo chain-of-states construction of transition paths}
\label{sec:sm_chain}
\label{sec:chain_mc}

To construct explicit transition paths between the two
configuration classes, we introduced an auxiliary chain of
configurations in the full configurational space. Each element of
the chain, hereafter referred to as an image, represents a complete
three-particle configuration,
\begin{equation}
X_m
=
\left\{
\left(\mathbf r_{m,i},\theta_{m,i}\right)
\right\}_{i=1}^{3},
\qquad
m=0,\ldots,M-1,
\label{eq:chain_image}
\end{equation}
where $m$ labels the position of the image along the chain and $i$
labels one of the three hard squares within that image. The images
are auxiliary variables and do not represent additional particles
or interacting physical replicas. Rather, the chain provides a
discrete representation of a candidate path in configurational
space connecting the two configuration classes.

The endpoint images, $X_0$ and $X_{M-1}$, were fixed to
representative maximally packed configurations of classes I and II,
respectively, and were uniformly rescaled to the packing fraction
at which the path was constructed. The initial chain contained
$M=400$ images. Because direct interpolation between the endpoints
generally produces overlapping intermediate configurations, the
internal images were initialized by duplicating the two valid
endpoint configurations. Specifically, each endpoint configuration
was copied 199 additional times, so that the initial chain contained
200 identical images in each endpoint state, separated by a single
interface. All images in the initial chain were therefore
overlap-free.

Neighboring images were coupled through the auxiliary harmonic
potential
\begin{equation}
U_{\rm spr}
=
\frac{k_{\rm spr}}{2}
\sum_{m=0}^{M-2}
\left[
d(X_m,X_{m+1})-\ell_0
\right]^2,
\label{eq:chain_spring}
\end{equation}
where $d(X_m,X_{m+1})$ is the distance between two complete
configurations, $k_{\rm spr}$ is the spring constant, and $\ell_0$
is the target spacing between neighboring images. These springs
are purely numerical restraints used to regularize the discretized
path and do not represent physical interactions between the hard
squares.

The configuration-space metric was defined after removing the
trivial symmetries associated with periodic wrapping, global
translation, particle relabeling, and the fourfold rotational
symmetry of a square. Consider two configurations
\begin{equation}
A
=
\left\{
\left(\mathbf r_i^A,\theta_i^A\right)
\right\}_{i=1}^{3},
\qquad
B
=
\left\{
\left(\mathbf r_i^B,\theta_i^B\right)
\right\}_{i=1}^{3},
\label{eq:two_configurations}
\end{equation}
in a square periodic box of side length L. For a given particle
permutation $\pi\in S_3$, we enumerated the periodic unwrapping
branches
\begin{equation}
\mathbf n_i
=
(n_{i,x},n_{i,y})
\in
\{-1,0,1\}^{2},
\end{equation}
and defined the corresponding unwrapped displacement of each
matched particle pair as
\begin{equation}
\Delta\mathbf r_i
=
\mathbf r_i^A
-
\mathbf r_{\pi(i)}^B
+
L\mathbf n_i.
\label{eq:unwrapped_displacement}
\end{equation}
For a fixed permutation and a fixed set of unwrapping branches,
the common translation that minimizes the total squared positional
mismatch is
\begin{equation}
\boldsymbol{\tau}_{\rm opt}
=
\frac{1}{3}
\sum_{i=1}^{3}
\Delta\mathbf r_i.
\label{eq:optimal_translation}
\end{equation}
The positional contribution to the configuration distance was
therefore defined as
\begin{equation}
D_{\rm pos}^2(A,B;\pi)
=
\frac{1}{3L^2}
\min_{\{\mathbf n_i\}}
\sum_{i=1}^{3}
\left|
\Delta\mathbf r_i
-
\boldsymbol{\tau}_{\rm opt}
\right|^2,
\label{eq:chain_positional_distance}
\end{equation}
where the minimum is taken over all combinations of the periodic
unwrapping branches. Equation~\eqref{eq:optimal_translation}
removes the global translational degree of freedom, leaving only
differences in the relative particle positions.

The orientational mismatch between two matched squares was reduced
modulo $\pi/2$,
\begin{equation}
\Delta\theta_{i,\pi(i)}
=
\theta_i^A-\theta_{\pi(i)}^B
-
\frac{\pi}{2}
\operatorname{round}
\left[
\frac{\theta_i^A-\theta_{\pi(i)}^B}{\pi/2}
\right].
\label{eq:chain_angular_difference}
\end{equation}
The total squared distance between configurations $A$ and $B$ was
then defined as
\begin{equation}
d^2(A,B)
=
\min_{\pi\in S_3}
\left[
D_{\rm pos}^2(A,B;\pi)
+
\frac{w_\theta}{3}
\sum_{i=1}^{3}
\Delta\theta_{i,\pi(i)}^2
\right].
\label{eq:chain_configuration_distance}
\end{equation}
This metric is invariant under periodic wrapping, global
translation, particle relabeling, and rotations of an individual
square by integer multiples of $\pi/2$.

The parameter $w_\theta$ sets the relative weighting of positional
and orientational differences. We used $w_\theta=0.01$ throughout.
The orientational term promotes smooth variations of the particle
orientations along the chain. Increasing $w_\theta$ made the
angular increments between neighboring images more uniform at the
expense of less uniform positional increments, but did not alter
the topology or qualitative geometry of the connecting path in our
tests. Thus, although the numerical arclength of the discretized
chain depends on the metric convention, the resulting overlap-free
transition channel was robust to moderate variations of
$w_\theta$.

The internal images were relaxed by Metropolis Monte Carlo. In each
trial move, an internal image $X_m$, with
$1\leq m\leq M-2$, and one of its three squares were selected at
random. The selected square was translated and rotated according to
\begin{equation}
\mathbf r_{m,i}
\rightarrow
\mathbf r_{m,i}+\delta\mathbf r,
\qquad
\theta_{m,i}
\rightarrow
\theta_{m,i}+\delta\theta,
\label{eq:chain_trial_move}
\end{equation}
while all other particles and images remained unchanged. The two
Cartesian components of $\delta\mathbf r$ and the angular
displacement $\delta\theta$ were drawn independently from
symmetric uniform distributions. Their proposal amplitudes were
adjusted together to maintain an acceptance rate of approximately
$0.3$.

A trial move that produced a hard-particle overlap under periodic
boundary conditions was rejected. If the updated image remained
overlap-free, only the two spring terms involving $X_m$ had to be
recalculated. The move was then accepted with probability
\begin{equation}
\alpha_{\rm acc}
=
\min
\left[
1,
\exp\left(-\beta\Delta U_{\rm spr}\right)
\right],
\label{eq:chain_acceptance}
\end{equation}
where $\Delta U_{\rm spr}$ is the change in the auxiliary spring
energy. We used reduced units in which $\beta =(k_{\rm B}T)^{-1}=1$.
During a single relaxation schedule, the spring constant was
increased monotonically from $10^{4}$ to $10^{13}$ in
multiplicative increments of $10^{0.5}$. At each fixed value of
$k_{\rm spr}$, we performed $10^{5}$ Monte Carlo sweeps before
proceeding to the next value. One sweep consisted of $3(M-2)$ randomly selected single-particle
trial moves, where $M$ denotes the current number of images in the
chain. Thus, each particle in each internal image was selected once
on average per sweep.
At small $k_{\rm spr}$, the weak spring bias allowed the internal
images to explore the overlap-free configurational space. As
$k_{\rm spr}$ was increased, the Metropolis acceptance criterion
progressively favored moves that reduced large separations between
neighboring images, thereby tightening and smoothing the chain.
We initially set $\ell_0=0$, so that the spring potential penalized
large neighboring-image separations without imposing a finite
target spacing. The value of $\ell_0$ was kept at zero for
$k_{\rm spr}<10^{10}$. For $k_{\rm spr}\geq10^{10}$, it was
gradually increased from $3.0\times10^{-4}$ to
$5.6\times10^{-4}$ and, at each stage, was chosen to be slightly
smaller than the minimum distance between neighboring images. The
finite target spacing reduced clustering and fluctuations in the
image distribution while preserving the resolution of the
discretized path. Moderate variations of $\ell_0$ within this
range affected the neighboring-image spacings but produced no
discernible change in the resulting transition path. The
relaxation was terminated after completion of the
$k_{\rm spr}=10^{13}$ stage.

During the relaxation, we monitored the neighboring-image
distances
\begin{equation}
d_m
=
d(X_m,X_{m+1}),
\qquad
m=0,\ldots,M-2.
\end{equation}
Because the two sides of the bottleneck generally have different
configurational arclengths, assigning equal numbers of images to
the two endpoint basins can produce different mean spacings on the
two portions of the chain. To obtain comparable resolution along
the full path, images were inserted into the portion with the
larger mean spacing and removed from the portion with the smaller
mean spacing. A new image was introduced by duplicating a
neighboring valid configuration, thereby preserving the
hard-particle constraint, whereas redundant internal images were
removed from oversampled portions of the path. After each insertion
or deletion, the affected neighboring distances and spring-energy
terms were recomputed, and the chain was relaxed again. This
procedure was repeated until the two sides of the bottleneck had
comparable mean image spacings. The endpoint images remained fixed
throughout. Using this Monte Carlo chain-of-states procedure, we obtained two distinct, densely discretized, overlap-free paths connecting the two configuration classes through the bottleneck, as shown in Fig.~3(f)--(i) and (k)--(n) of the main text.

\section{Geometric origin of the configurational-space separation density}
\label{sec:sm_geometry}

\begin{figure}
    \centering \includegraphics[width=2.7 in]{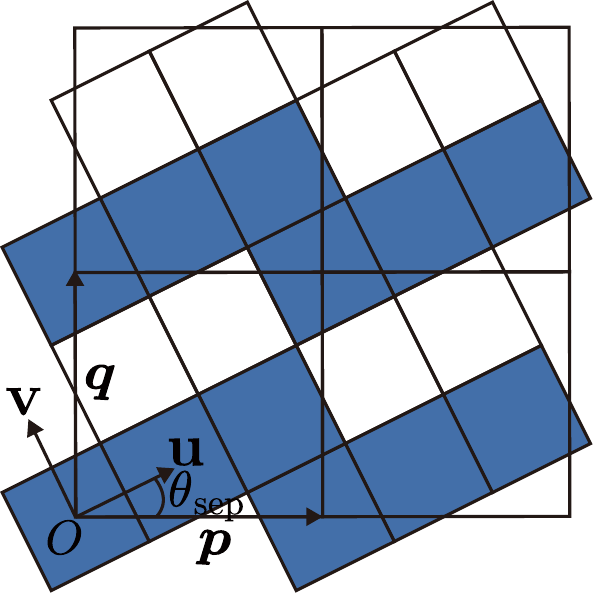}
    \caption{Geometric construction of the tilted bottleneck configuration at
$\phi=0.6$. The box vectors $\mathbf{p}$ and $\mathbf{q}$ are
constructed from the square-edge directions $\mathbf{u}$ and
$\mathbf{v}$ according to Eq.~(\ref{eq:box_vectors}), with
$(m,n)=(2,1)$.}
    \label{fig_square}
\end{figure}

Here we provide a geometric interpretation of the configurational-space
separation density $\phi_{\rm sep}=0.6$. The argument is motivated by
the transition pathways shown in Fig.~3 of the main text. Along all
three pathways, the system passes through a narrow bottleneck in which
the three squares are nearly parallel and nearly collinear. Near this
bottleneck, their rotations are strongly constrained, and the remaining
configurational freedom is dominated by sliding along their common edge
directions.

Let $a$ be the side length of each square and $L$ the side length of the square periodic box. The packing fraction is
\begin{equation}
    \phi=\frac{3a^2}{L^2}.
\label{eq:packing_fraction_geometry}
\end{equation}
To identify the characteristic box size associated with the
bottleneck, we approximate the limiting configuration by three
parallel squares that just fit in the periodic box without overlap.
The corresponding geometric construction is illustrated in
Fig.~\ref{fig_square}. Let $\mathbf{u}$ and $\mathbf{v}$ be two
orthonormal directions parallel to the square edges. For a square
periodic box, the two box vectors must have equal length and be
orthogonal. If they are constructed from integer steps along
$\mathbf{u}$ and $\mathbf{v}$, they can be written as
\begin{equation}
    \mathbf{p}=a(m\mathbf{u}-n\mathbf{v}),
    \qquad
    \mathbf{q}=a(n\mathbf{u}+m\mathbf{v}),
\label{eq:box_vectors}
\end{equation}
where $m$ and $n$ are non-negative integers. These vectors satisfy
$|\mathbf{p}|=|\mathbf{q}|=a\sqrt{m^2+n^2}$ and
$\mathbf{p}\cdot\mathbf{q}=0$. Hence,
\begin{equation}
    L=a\sqrt{m^2+n^2},
    \qquad
    \phi=\frac{3}{m^2+n^2}.
\label{eq:integer_geometry}
\end{equation}
The smallest integer constructions are listed in
Table~\ref{tab:parallel_cases}. Cases with $n=0$ correspond to axial
arrangements in which the square edges are aligned with the box
directions. The smaller tilted construction $(m,n)=(1,1)$ cannot
accommodate three squares. The smallest physically allowed tilted
construction is therefore $(m,n)=(2,1)$, which gives
\begin{equation}
    L_{\rm sep}=\sqrt{5}a,
    \qquad
    \phi_{\rm sep}=\frac{3}{5}=0.6.
\label{eq:separation_density_geometry}
\end{equation}
For the $(m,n)=(2,1)$ geometry, one may choose
\begin{equation}
    \mathbf{u}=\frac{1}{\sqrt{5}}(2,1),
    \qquad
    \mathbf{v}=\frac{1}{\sqrt{5}}(-1,2).
\label{eq:bottleneck_basis}
\end{equation}
The corresponding orientation of the squares relative to the box is
\begin{equation}
\theta_{\rm sep}
= \tan^{-1}\left(\frac{1}{2}\right)
\approx 26.6^\circ,
\label{eq:separation_angle}
\end{equation}
which is the angle indicated in Fig.~3 of the main text.

This construction provides a geometric picture of the connecting
channel near the separation density. As $\phi$ approaches
$\phi_{\rm sep}$ from below, the channel becomes increasingly narrow,
making direct transitions between the two configurational regions
progressively more difficult to observe by MCMC. At
$\phi=\phi_{\rm sep}$, the tilted bottleneck becomes a marginal,
just-fitting configuration: the three squares retain only translational
degrees of freedom, corresponding to sliding along their common edge
directions, whereas any rotation away from this limiting configuration
would produce overlap. This explains why direct transitions between the two configurational regions are unlikely to be observed in MCMC simulations at this density.
For $\phi>\phi_{\rm sep}$, or equivalently
$L<\sqrt{5}a$, the corresponding nearly parallel tilted arrangement
cannot fit in the periodic box without overlap.

We emphasize that this construction does not constitute a rigorous
proof that all possible transition pathways are forbidden for
$\phi>\phi_{\rm sep}$, which would require exhausting all possible
relative orientations, contact geometries, and periodic-image
relations. Rather, it provides a geometric explanation of the
bottleneck identified in the simulations and is consistent with both
the pressure-branch separation and the transition pathways.

\begin{table}
\centering
\caption{
Smallest integer constructions of parallel squares compatible with a
square periodic box. The box side length is
$L=a\sqrt{m^2+n^2}$, and the corresponding packing fraction for three
squares is $\phi=3/(m^2+n^2)$.
}
\label{tab:parallel_cases}
\setlength{\tabcolsep}{14pt}
\begin{tabular}{c c c l}
\hline
$(m,n)$ & $L^2/a^2$ & $\phi$ & Interpretation \\
\hline
$(1,0)$ & $1$  & $3$    & Too small to contain three squares \\
$(1,1)$ & $2$  & $3/2$  & Too small to contain three squares \\
$(2,0)$ & $4$  & $3/4$  & Axial arrangement \\
$(2,1)$ & $5$  & $3/5$  & Smallest tilted arrangement \\
$(2,2)$ & $8$  & $3/8$  & Tilted arrangement at lower packing fraction \\
$(3,0)$ & $9$  & $1/3$  & Axial arrangement at lower packing fraction \\
$(3,1)$ & $10$ & $3/10$ & Tilted arrangement at lower packing fraction \\
\hline
\end{tabular}
\end{table}

\section{Classification of high-density configurations}
\label{sec:sm_classification}

At the configurational-space separation density
$\phi_{\rm sep}=0.6$, the connecting channel between the two
configurational regions narrows to a marginal bottleneck, as discussed
in Sec.~\ref{sec:sm_geometry}. In the MCMC trajectories used here, the two sides of this
bottleneck remain dynamically separated and correspond to the two
pressure branches identified in the main text. We therefore denote
the configurations sampled on the two sides as classes I and II and
use them to construct labeled data. Specifically, we performed MCMC
simulations initialized in representative configurations of the two
classes at $\phi=0.6$, using the same MCMC protocol as described in Sec.~II.
We collected $10^7$ configurations for each class,
yielding a total of $2\times10^7$ labeled configurations.

Because a square has fourfold rotational symmetry, orientations differing by integer multiples of $\pi/2$ are physically equivalent. In configurations generated by uniform insertion sampling, the particle orientations span the full interval $[0,2\pi)$, whereas within a finite MCMC trajectory the particles may not explore all symmetry-equivalent orientations because their rotations are constrained by excluded-volume interactions with the other particles. We therefore reduced the orientation of each square modulo $\pi/2$ and mapped it onto $[0,\pi/2)$. In addition, each training configuration was subjected to a random global translation, applied identically to all three particles, and to a random permutation of the particle labels. These operations reduce the dependence of the classifier on the arbitrary choice of coordinate origin and particle labeling.

We used a multilayer perceptron (MLP) to distinguish the two configurational classes. The raw particle data $(x,y,z,q_w,q_x,q_y,q_z)$ were converted into a nine-dimensional input vector $(x_i,y_i,\theta_i)_{i=1}^{3}$, where $\theta_i$ is the in-plane orientation obtained from the quaternion and reduced modulo $\pi/2$. The particle coordinates were normalized by the box half-length and wrapped into $[-1,1)$ according to the periodic boundary conditions. The fully connected network contained three hidden layers with 200, 40, and 5 neurons, respectively, each using a ReLU activation function, followed by a single sigmoid output unit representing the predicted probability $p_{\mathrm{II}}$ of belonging to class II. The network was trained for 200 epochs using the Adam optimizer and binary cross-entropy loss, with a batch size of 65536 and an 80\%/20\% training/test split. No misclassifications were observed among the $4.0\times10^6$ configurations in the held-out test set. Configurations with $p_{\mathrm{II}}\geq0.5$ were assigned to class II, and the remaining configurations were assigned to class I.

For packing fractions $\phi'>0.6$, the allowed configurational
space expressed in fractional coordinates satisfies
\[
\Omega(\phi')\subseteq\Omega(0.6),
\]
because increasing the packing fraction can only remove
configurations that develop particle overlaps. For any allowed
configuration at $\phi'>0.6$, the same wrapped fractional
coordinates therefore define an allowed configuration at
$\phi=0.6$. Since the box is larger at $\phi=0.6$, this
corresponds physically to a dilated representation of the
higher-density configuration, in which the interparticle
separations are increased while the fractional coordinates and
particle orientations remain unchanged. We assign the class of a higher-density configuration according to
which side of the $\phi=0.6$ bottleneck contains its dilated
representation. These two sides are operationally defined by the
class-I and class-II MCMC ensembles used to train the classifier.
Because increasing the packing fraction can only remove allowed
configurations, it cannot create a new connecting channel between
the two sides. The class-I/class-II labeling at $\phi=0.6$ therefore
provides a consistent continuation to higher packing fractions.
Moreover, a higher-density configuration and its dilated
representation have identical fractional coordinates and particle
orientations, and hence correspond to the same input representation
used by the classifier. We therefore apply the classifier trained at
$\phi=0.6$ directly to higher-density configurations using the same
coordinate and orientation preprocessing.

As a consistency check, we applied the trained classifier to the
$10^7$ configurations generated by uniform insertion sampling at
$\phi=0.6$ and evaluated the pressure separately for configurations
assigned to classes I and II. Using the resulting class-resolved
pressures, together with the ensemble-averaged pressure
$P_{\mathrm{tot}}$, in the relation derived above, we obtain
\[
    \frac{Q_2}{Q_{\mathrm{tot}}} \approx 0.0503,
\]
in close agreement with the value $0.0508$ obtained from the two MCMC
pressure branches. This agreement provides a consistency check on the
two-region description and the configurational classification.

We further applied the trained classifier to configurations generated
by uniform insertion sampling at other packing fractions and evaluated
the pressure separately for configurations assigned to classes I and
II. The resulting class-resolved pressures are shown by the black lines
in Fig.~2(c) of the main text.

\section{Order parameter in the dilute limit}
\label{sec:sm_dilute}

\begin{figure}
    \centering
    \includegraphics[width=5 in]{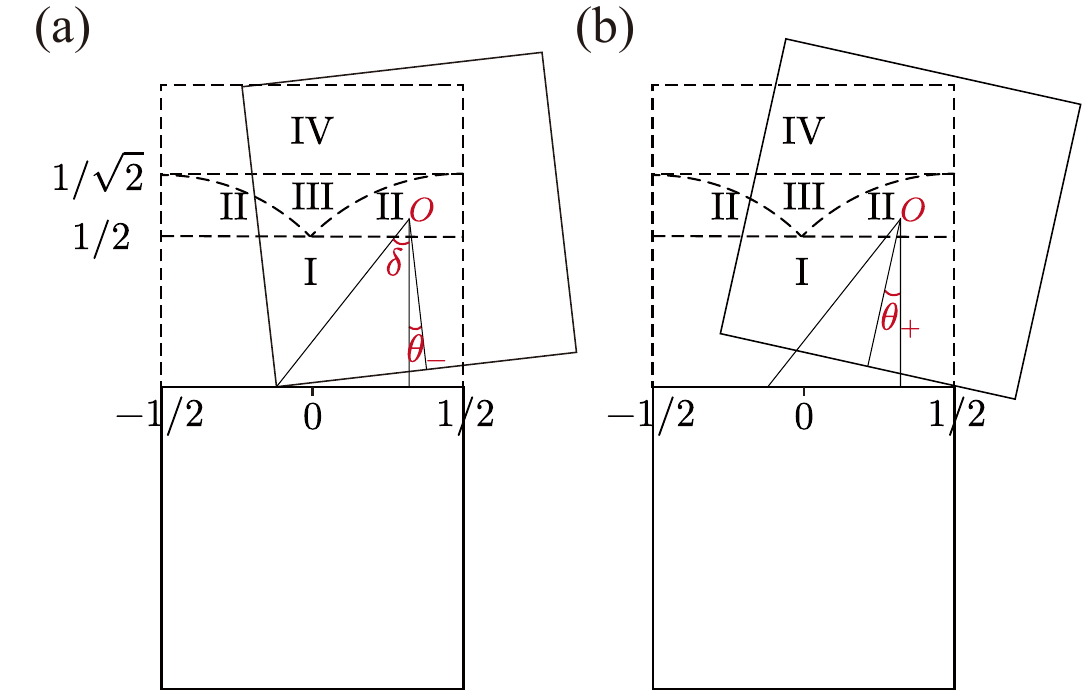}
    \caption{Geometric construction of the limiting rotations of the second square
when its center lies in region II. The lower square is the reference square, and the upper square is the second square. Panels (a) and (b) illustrate the counterclockwise and clockwise limiting rotations, $\theta_{-}$ and $\theta_{+}$, respectively, at which the two squares first come into contact. The dashed lines partition the possible locations of the center $O$ of the second square into regions I--IV. The auxiliary angle $\delta$ is indicated.}
    \label{fig_dilute}
\end{figure}

In the dilute limit, correlations beyond the pair excluded-volume
constraint can be neglected, so that the distribution of the local
order parameter is determined by the geometry of two hard squares.
We fix one square as the reference and consider a second square whose
center lies in one of the four adjacent square regions shown in
Fig.~\ref{fig_dilute}. Both coordinates of the second-square center
are measured in units of the square side length $a$. We denote its
lateral displacement from the midpoint of the corresponding side of
the reference square by $s=d/a\in[-1/2,1/2]$, and its perpendicular
distance from that side by $h\in[0,1]$.

Because a square has fourfold rotational symmetry, it is sufficient
to consider one angular period of length $\pi/2$. We take the
orientation parallel to the reference square as $\alpha=0$ and
restrict the relative orientation to
$\alpha\in[-\pi/4,\pi/4]$, with positive $\alpha$ corresponding to
clockwise rotation. For fixed $(s,h)$, the statistical weight in the
dilute limit is proportional to the total angular range over which
the two squares remain non-overlapping. We denote this allowed
angular measure by $\mathcal{A}(s,h)$.

The geometry is symmetric under $s\rightarrow-s$, so we introduce
$\xi=|s|$. The normalized dilute-limit reference curve can then be
written as
\begin{equation}
f_{\rm id}(s)=\frac{F(|s|)}{F(0)},
\qquad
F(\xi)=\int_0^1 \mathcal{A}(\xi,h)\,\mathrm{d}h .
\label{eq:dilute_F}
\end{equation}
The normalization is chosen such that $f_{\rm id}(0)=1$, consistent
with that used in Fig.~4(c) of the main text.

The dashed construction in Fig.~\ref{fig_dilute} divides the possible
locations of the second-square center into regions I--IV. In region I,
$0\leq h<1/2$, the two squares overlap for all orientations. In region IV,
$1/\sqrt{2}\leq h \leq 1$, the two squares remain non-overlapping over the
entire angular interval, giving $\mathcal{A}(\xi,h)=\pi/2$. 

We next consider region II, where
$1/2\leq h<h_c(\xi)$. The clockwise and counterclockwise rotations are
limited by different contact geometries. We define the auxiliary angle
\begin{equation}
\delta(h)=
\arccos\left(\frac{h}{\sqrt{2}/2}\right).
\label{eq:dilute_delta}
\end{equation}
As illustrated in Fig.~\ref{fig_dilute}, we denote the maximum
counterclockwise and clockwise rotation angles by $\theta_{-}$ and
$\theta_{+}$, corresponding to panels (a) and (b), respectively. The
counterclockwise limit is
\begin{equation}
\theta_{-}(h)=\frac{\pi}{4}-\delta(h),
\label{eq:dilute_theta_minus}
\end{equation}
whereas the clockwise limit is
\begin{equation}
\begin{split}
\theta_{+}(\xi,h)
&=
\frac{\pi}{2}
-\arctan\left(\frac{1-2\xi}{2h}\right)
-\arcsin\left[
\frac{1}{\sqrt{4h^2+(1-2\xi)^2}}
\right].
\end{split}
\label{eq:dilute_theta_plus}
\end{equation}
Since positive $\alpha$ corresponds to clockwise rotation, the allowed
orientations in region II satisfy
$-\theta_{-}(h)\leq\alpha\leq\theta_{+}(\xi,h)$, and the corresponding
angular measure is $\theta_{+}(\xi,h)+\theta_{-}(h)$.
The boundary between regions II and III is reached when, upon
clockwise rotation, the lower-right vertex of the second square just
touches the upper-right vertex of the reference square. The
corresponding crossover height is
\begin{equation}
h_c(\xi)=
\sqrt{
\left(\frac{\sqrt{2}}{2}\right)^2
-\left(\xi-\frac{1}{2}\right)^2
}.
\label{eq:dilute_hc}
\end{equation}
At $h=h_c(\xi)$, the clockwise and counterclockwise limiting angles
become equal.

For $h_c(\xi)\leq h<1/\sqrt{2}$, corresponding to region III, the
limiting contacts in both rotational directions are geometrically
equivalent: in each case, a vertex of the second square comes into
contact with the upper side of the reference square. The clockwise
and counterclockwise limiting angles therefore have the same
magnitude $\theta_{-}(h)$, and the allowed angular measure is
$2\theta_{-}(h)$.

Combining the four geometric regions, the allowed angular measure is
\begin{equation}
\mathcal{A}(\xi,h)=
\begin{cases}
0,
& 0\leq h<\dfrac{1}{2},
\\
\theta_{+}(\xi,h)+\theta_{-}(h),
& \dfrac{1}{2}\leq h<h_c(\xi),
\\
2\theta_{-}(h),
& h_c(\xi)\leq h<\dfrac{1}{\sqrt{2}},
\\
\dfrac{\pi}{2},
& \dfrac{1}{\sqrt{2}}\leq h \leq 1.
\end{cases}
\label{eq:dilute_angular_measure}
\end{equation}
The four cases correspond to regions I--IV, respectively. Substituting Eq.~\eqref{eq:dilute_angular_measure} into
Eq.~\eqref{eq:dilute_F}, we obtain
\begin{equation}
\begin{split}
F(\xi)
&=
\int_{1/2}^{h_c(\xi)}
\left[
\theta_{+}(\xi,h)+\theta_{-}(h)
\right]\mathrm{d}h
+\int_{h_c(\xi)}^{1/\sqrt{2}}
2\theta_{-}(h)\,\mathrm{d}h
+\int_{1/\sqrt{2}}^{1}
\frac{\pi}{2}\,\mathrm{d}h .
\end{split}
\label{eq:dilute_F_explicit}
\end{equation}
The dilute-limit reference curve shown in
Fig.~4(c) of the main text is therefore
\begin{equation}
f_{\rm id}\left(\frac{d}{a}\right)
=
\frac{F\!\left(\left|d/a\right|\right)}{F(0)}.
\label{eq:dilute_final}
\end{equation}
Numerical evaluation of these integrals yields the gray dashed curve
shown in Fig.~4(c) of the main text.

\bibliography{ref}